\documentclass{nature}
\usepackage{graphicx}
\makeatletter
\let\saved@includegraphics\includegraphics
\AtBeginDocument{\let\includegraphics\saved@includegraphics}
\renewenvironment*{figure}{\@float{figure}}{\end@float}
\makeatother

\usepackage{multirow}
\usepackage{subfigure}
\usepackage{mathtools}
\usepackage[colorlinks=true,linkcolor=blue,citecolor=black,urlcolor=blue]{hyperref}
\usepackage{authblk}
\usepackage[labelfont=bf]{caption}
\usepackage[figurename=Figure]{caption}
\usepackage{siunitx}
\usepackage[dvipsnames]{xcolor}
\usepackage{soul}
\usepackage{xfrac}
\usepackage{amsmath}
\usepackage{amssymb}
\usepackage{soul,xcolor}
\usepackage{lineno}
\usepackage{braket}

\usepackage{amsfonts}
\usepackage{booktabs}
\usepackage{tabularx}
\usepackage{adjustbox}

\graphicspath{{./figures/}}
\newcommand{\iu}{\mathrm{i}}
\newcommand{\VEC}[1]{\mathbf{#1}}

\newcommand{\new}[1]{{\color{black} #1}}

\begin{document}

\setstcolor{red}

\title{Triple-Q state in magnetic breathing kagome lattice}

\author[1,2,3,4,$\dag$]{Hangyu Zhou}

\author[5]{Manuel dos Santos Dias}
\author[6]{Shijian Bao}
\author[6]{Hanchen Lu}
\author[3]{Youguang Zhang}
\author[1,$\ddagger$]{Weisheng Zhao}
\author[2,7,*]{Samir Lounis}

\affil[1]{Fert Beijing Institute, School of Integrated Circuit Science and Engineering, Beihang University, Beijing 100191, China.}
\affil[2]{Peter Gr\"unberg Institut, Forschungszentrum J\"ulich \& JARA, 52425 J\"ulich, Germany}
\affil[3]{School of Electronic and Information Engineering, Beihang University, Beijing 100191, China}
\affil[4]{Shenyuan Honors College, Beihang University, Beijing 100191, China}
\affil[5]{Scientific Computing Department, STFC Daresbury Laboratory, Warrington WA4 4AD, United Kingdom}
\affil[6]{China International Engineering Consulting Corporation, Beijing 100048, China
}
\affil[7]{Institut für Physik, Martin Luther University Halle-Wittenberg, 06099 Halle, Germany}
\affil[$\dag$]{hangyu.zhou@buaa.edu.cn}
\affil[$\ddagger$]{weisheng.zhao@buaa.edu.cn}
\affil[*]{samir.lounis@physik.uni-halle.de}

\maketitle

\begin{abstract}
Magnetic frustration in two-dimensional spin lattices with triangular motifs underpins a series of exotic states, ranging from multi-Q configurations to disordered spin-glasses. 
The antiferromagnetic kagome lattice, characterized by its network of corner-sharing triangles, represents a paradigmatic frustrated system exhibiting macroscopic degeneracy.
Expanding upon the kagomerization mechanism, 
we  focus on the magnetic breathing kagome lattice formed by a Mn monolayer deposited on a heavy metal substrate and capped with \textit{h}-BN. 
The Mn kagome arrangement induces pronounced magnetic frustration, as evidenced by the nearly flat bands derived from spin spiral energy calculations. 
Including further-neighbor interactions reveals a spin spiral energy minimum along the $\mathbf{\Gamma}$-K line and an intriguing triple-Q state with nonzero topological charge, potentially leading to highly nonlinear Hall effects.
Furthermore, the flat band properties can further give rise to an even more complex spin configuration, marked by several Q-pockets in the spin structure factor.
These results present a fertile ground for advancing the study of multi-Q states and exploring emergent topological phenomena.
\end{abstract}

\section*{Introduction}
 Frustration in magnetic systems has attracted great interest due to the ability to support diverse unconventional states, such as spin-glass behaviour~\cite{PhysRevB.81.014406}, incommensurate magnetic order~\cite{doi:10.1143/JPSJ.79.011003,PhysRevB.81.094407} and noncoplanar magnetic patterns~\cite{PhysRevB.103.054422,doi:10.1126/sciadv.aau3402}.
 Magnetic frustration can arise from the geometry of the lattice or competing exchange interactions. The antiferromagnetic (AFM) Heisenberg model on the two-dimensional (2D) hexagonal lattice is a paradigm for frustrated magnets. With only AFM nearest-neighbor (n.n.) interactions, the ground state of this system is the three-sublattice N\'eel structure consisting of coplanar spins forming $\pm 120^\circ$ angles between nearest neighbors, which is commensurate to the underlying lattice. Including further neighbors or even higher-order spin interactions, more complex magnetic states such as superposition of spin spirals~\cite{PhysRevLett.86.1106,PhysRevLett.124.227203,PhysRevB.95.224424} or atomic scale spin lattices~\cite{Heinze2011,PhysRevB.92.020401} are stabilized. Those multi-Q states are intriguing, as they often exhibit nonzero vector chirality, $\VEC{S}_i\times\VEC{S}_j$, and scalar chirality, $\VEC{S}_{i}\cdot(\VEC{S}_j\times\VEC{S}_k)$. These chiralities lead to unconventional phenomena, such as the spin Hall effect~\cite{PhysRevLett.95.057205,PhysRevLett.113.196602}, the topological Hall effect~\cite{PhysRevB.45.13544,PhysRevLett.83.3737,PhysRevB.62.R6065,Surgers2014,Xu2024}, the noncollinear Hall effect ~\cite{PhysRevLett.126.147203}, nonreciprocal transport~\cite{Ishizuka2020,PhysRevB.101.220403},  complex magnetoresistances~\cite{Crum2015,LimaFernandes2022} and the formation of topological orbital moments ~\cite{dosSantosDias2016,PhysRevB.94.121114,PhysRevB.98.125420,PhysRevB.98.094428} 
 as well as topological high-order interactions~\cite{Grytsiuk2020}. 
 A notable example of a multi-Q state is the triple-Q state~\cite{PhysRevLett.86.1106}, characterized by the superposition of three symmetry-equivalent spin spirals,  where the relative angle between all nearest-neighbor spins is the tetrahedron angle of 109.47$^\circ$~\cite{WORTMANN200257}.  
 The triple-Q state \new{in 2D} was first predicted in an Mn monolayer on the Cu(111) substrate and then experimentally  observed in an hcp-stacked Mn monolayer on Re(0001)\cite{PhysRevLett.124.227203}. 
Subsequently, the triple-Q state has also been observed in Pd/Mn and Rh/Mn bilayers on Re(0001) through SP-STM studies~\cite{PhysRevB.108.L180411}

In contrast to the 2D hexagonal lattice, the classic ground state of Heisenberg kagome antiferromagnets can exhibit macroscopical degeneracy, with a set of coplanar states being selected by the order-by-disorder mechanism~\cite{PhysRevB.45.7536,PhysRevB.78.094423}. For quantum spins on the kagome lattice, exotic ground states such as spin liquids or valence-bond crystal states may be stable~\cite{doi:10.1126/science.1201080,PhysRevB.83.212401,Jiang2012,PhysRevB.76.180407}.
Mn is a good candidate for investigating magnetic frustration as it often forms lattices composed of triangles and tends to have AFM interactions.
In this work, we explore the magnetic breathing kagome lattice formed by a Mn monolayer deposited on a heavy metal substrate and capped by \textit{h}-BN that is made possible by the kagomerization mechanism uncovered in our prior study~\cite{Zhou2024}.
The magnetic properties and magnetic interactions, including Heisenberg exchange interactions and Dzyaloshinskii-Moriya interactions (DMI), have been investigated. Compared to the triangular lattice of Mn on Pt substrate, the kagome structure of Mn induces strong frustration, as evidenced by the almost flat bands obtained for spin spiral energies not only when considering the n.n. interactions but also when including further-neighbor interactions.
Due to the microscopic properties of this magnetic breathing kagome lattice, the spin spiral energy minimum is found on the $\Gamma$-K line, and a triple-Q state is uncovered through atomistic spin dynamics.
The strong noncoplanarity of this spin texture is characterized via its structure factor and by its non-vanishing topological charge density.
Additionally, a more complex spin state, characterized by several Q-pockets in the spin structure factor, emerges due to the flat bands.
Such spin textures enabled by the kagomerization mechanism may offer a platform for the investigation of novel topological phenomena.

\section*{Results}
\begin{figure}
    \centering
    \includegraphics[width=1\linewidth]{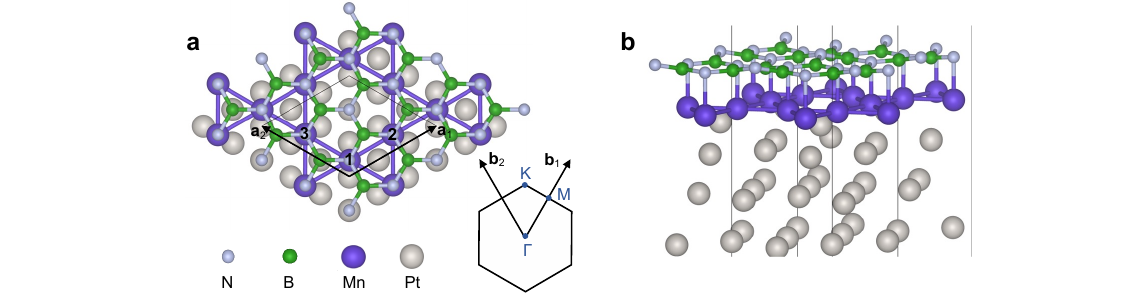}
    \caption{Atomic structure of Pt/Mn/\textit{h}-BN. \textbf{a} Top view with the hexagonal 2D Brillouin zone. $\VEC{a}_1$ and $\VEC{a}_2$ are the in-plane Bravais vectors and $a$ is the lattice parameter of the surface unit cell. $\VEC{b}_1$ and $\VEC{b}_2$ are the vectors of the corresponding  reciprocal lattice. \textbf{b} Side view. In each unit cell, there are three Mn atoms, labelled as Mn(1), Mn(2) and Mn(3).}
    \label{fig-1}
\end{figure}

We show in Fig.~\ref{fig-1} the atomic structure of the considered kagome structure Pt/Mn/\textit{h}-BN.
As we found in Ref.~\cite{Zhou2024}, this metastable kagome arrangement of Mn atoms is enabled by the \textit{h}-BN overlayer and should be experimentally realizable under appropriate growth conditions.
In the optimised structure of  Pt/Mn/\textit{h}-BN, all N atoms of the unit cell are on top of a Mn atom, except for one N that instead attracts one Pt atom from the substrate towards the hexagonal hole of the kagome lattice. More details on how the kagome lattice is formed can be found in our previous work~\cite{Zhou2024}. 

\begin{figure}
    \centering
    \includegraphics[width=1\linewidth]{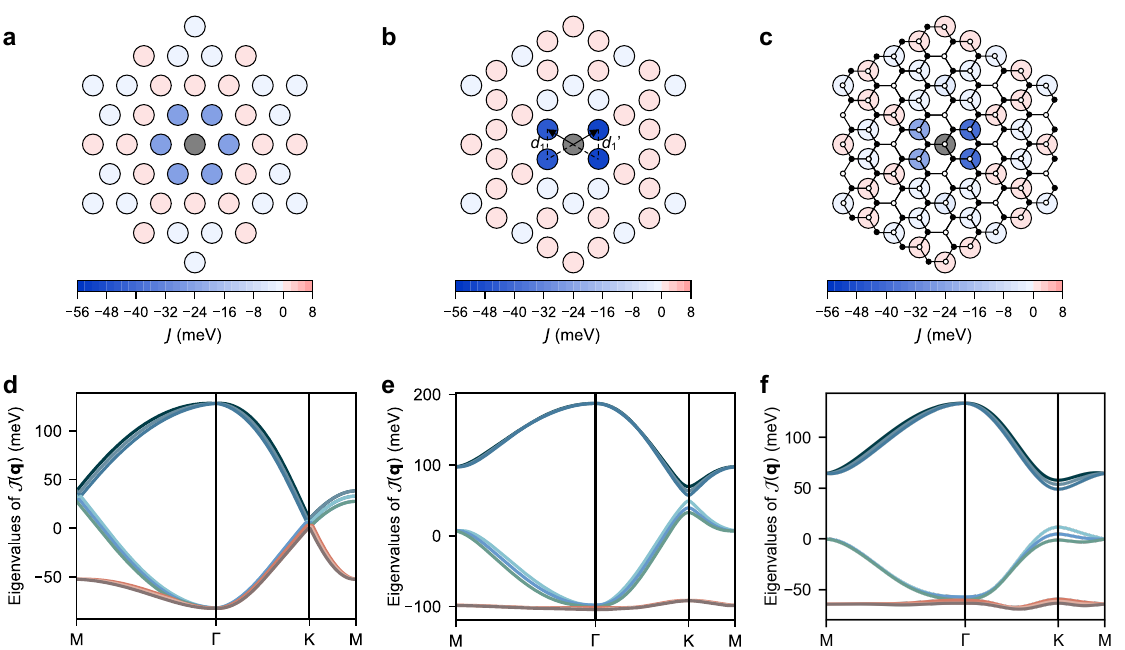}
    \caption{Heisenberg exchange interactions and their lattice Fourier transform. \textbf{a}-\textbf{c} Spatial distribution of the Heisenberg exchange interactions for: \textbf{a} Pt/Mn with Mn in triangular lattice; \textbf{b} Pt/Mn with Mn in kagome lattice; \textbf{c} Pt/Mn/\textit{h}-BN kagome structure. In panel \textbf{a}, $d_1$=2.46\ \AA\ ($d_1'$=2.41\ \AA) is the side length of the large (small) equilateral triangle in the kagome structure. The small white and black circles in panel \textbf{c} represent N and B atoms, respectively. \textbf{d}-\textbf{f} Eigenvalues of the Fourier-transformed magnetic interactions, including Heisenberg exchange interactions and DMI for: \textbf{a} Pt/Mn hexagonal structure; \textbf{b} Pt/Mn kagome structure; \textbf{c} Pt/Mn/\textit{h}-BN kagome structure.}
    \label{fig-2}
\end{figure}

To analyse the source of the magnetic frustration, we calculated the pairwise interactions, including Heisenberg exchange ($J$) and DM vectors ($\VEC{D}$), for the Pt/Mn/\textit{h}-BN kagome and Pt/Mn 2D hexagonal structures. The magnetic interactions were calculated from the disordered local moment (DLM) state (see Methods), which serves as an excellent reference for the magnetic state due to its neutrality towards both ferromagnetic (FM) and AFM preferences.
In order to assist the analysis, we also removed the \textit{h}-BN overlayer from the Pt/Mn/\textit{h}-BN kagome structure while retaining the kagome arrangement of the Mn atoms.
We present the spatial distribution of the Heisenberg exchange interactions $J$ of the three structures in Fig.~\ref{fig-2}a-c. Each circle represents a Mn atom and is colored as a function of $J$ values with respect to the central Mn (gray circle) which is taken as a reference. The value of $J$ is given by the color scale, with red (blue) for FM (AFM) coupling. Figure~\ref{fig-2}a depicts the results obtained for the Pt/Mn triangular structure with $C_{\rm 3v}$ symmetry. The n.n. distance is $d_1=2.81$\ \AA, the spin moment of the Mn atoms is 4.20 $\mu_{\rm B}$ and the strong AFM interactions in the six nearest neighbours lead to the 120$^\circ$ N\'eel state.
In the kagome Pt/Mn structure shown in Fig.~\ref{fig-2}b, the n.n. bonds are shorter and unequal, so that four nearest neighbors form one large and one small equilateral triangle, with side lengths of $d_1=2.46$\ \AA\ and $d_1'=2.41$\ \AA, respectively.
The spin moment is reduced to 3.66 $\mu_{\rm B}$, and  the slightly shorter distance of $d_1'$ compared to $d_1$ leads to a stronger AFM exchange interaction, with a difference of 7 meV. With the presence of \textit{h}-BN, as  illustrated in Fig.~\ref{fig-2}c, the magnetic moment of Mn decreases further to 2.15 $\mu_{\rm B}$ and the asymmetry in the exchange interactions increased to approximately 16 meV. Such a significant asymmetry in the nearest-neighbor $J$ values motivates us to term it a magnetic breathing kagome lattice, with $J_1/J_1'=1.6$ measuring the strength of the breathing bond alternation~\cite{PhysRevB.105.L100407}.

In Fig.~\ref{fig-2}d-f, we present the eigenvalues of the Fourier-transformed magnetic interactions $\mathcal{J}(\VEC{q})$ as a function of reciprocal momentum vector $\VEC{q}$, including Heisenberg exchange and DMI with a converged real-space cut off radius of 5$a$, where $a =4.87${\ \AA} is the lattice constant of Pt, as defined in the Methods.
The eigenvector associated with the lowest energy eigenvalue is a proxy for the magnetic ground state, and to explore the energy landscape we plot the eigenvalues along high-symmetry lines in the 2D Brillouin zone.
 The Heisenberg exchange interactions control the overall dispersion of the eigenvalues of $\mathcal{J}(\VEC{q})$, while DMI lifts their degeneracy.
In Fig.~\ref{fig-2}d, the global minimum is found at the $\Gamma$ point of the Pt/Mn triangular lattice and corresponds to the N\'eel state --- AFM merons~\cite{doi:10.1021/acs.jpclett.3c02419} can be found in this N\'eel background (see Supplementary Fig. S1).
In contrast to the triangular lattice, the eigenvalues of $\mathcal{J}(\VEC{q})$ for the Pt/Mn kagome structure shown in Fig.~\ref{fig-2}e have a rather flat band at -100 meV, which is due to the geometric frustration built into the kagome lattice.
Such flat bands can lead to the quasi-degeneracy of multiple magnetic states, increasing the importance of quantum and thermal fluctuations.
The asymmetry of the n.n.\ Heisenberg exchange interactions shown in Fig.~\ref{fig-2}b leads to the gap opening at the K point seen in Fig.~\ref{fig-2}e. 
The presence of \textit{h}-BN increases the size of the gap at the K point due to the larger asymmetry in the n.n.\ $J$ (Fig.~\ref{fig-2}c), and modifies the dispersion of the flat bands with the energy minimum at $\VEC{q}=(0.195,0.195)$ (in reciprocal lattice units), as shown in Fig.~\ref{fig-2}f.

To further understand the properties of the flat band and the origin of the energy minimum we have repeated this analysis considering different real-space cutoff radii in the calculation of $\mathcal{J}(\VEC{q})$, see Supplementary Fig.~S2.
We found that the flat bands persist even when considering only the highly asymmetric n.n.\ interactions, and that the energy minimum is essentially determined by the long-range Heisenberg exchange interactions, with the DMI making a small correction.
We conclude that unconventional phenomena related to flat bands can also be expected in this type of magnetic breathing kagome lattice.

\begin{figure}
    \centering
    \includegraphics[width=1\linewidth]{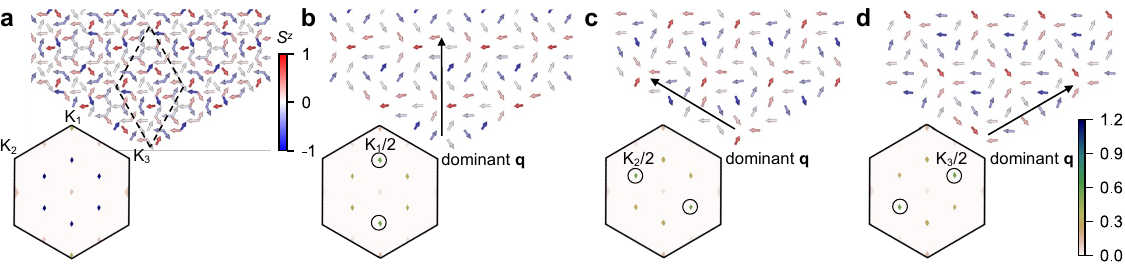}
    \caption{\new{The triple-Q state and its spin structure factor.}
    \textbf{a} The triple-Q state for the Pt/Mn/\textit{h}-BN kagome structure.
    The spins are colored as shown in  the colorbar: red is $+z$, blue is $-z$, and gray is in-plane.
    The dashed rhombus indicates the magnetic unit cell.
    The intensity of its spin structure factor is show as a colormap on the 2D Brillouin zone, where the darker the color, the higher the intensity.
    The triple-Q state can be decomposed into three sublattices: \textbf{b} sublattice of Mn(1); \textbf{c} sublattice of Mn(2); \textbf{d} sublattice of Mn(3).
    Insets in \textbf{b}-\textbf{d} are the structure factor of each sublattice.
    The main $\VEC{q}$ direction of each sublattice is circled in black, marked as K\textsubscript{1}/2, K\textsubscript{2}/2 or K\textsubscript{3}/2.
    The atomistic spin dynamics simulations were performed on a 30$\times$30 supercell in the zero temperature limit.}
    \label{fig-3}
\end{figure}

\new{The information obtained from the $\mathcal{J}(\VEC{q})$ was used to inform and interpret atomistic spin dynamics simulations to fully characterize the magnetic ground state.}
We found a highly noncollinear state in Fig.~\ref{fig-3}a with a magnetic unit cell comprising 36 sites (dashed  rhombus).
To gain more insights, we plot next to the magnetic structure the intensity of the spin structure factor $S(q) = \sqrt{\VEC{S}(\VEC{q})^*\cdot\VEC{S}(\VEC{q})}$ (see Methods).
The six dark blue spots in the 2D Brillouin zone show that this is a triple-Q state with the dominant $\VEC{q}$ at the K/2 points.
However, the periodicity revealed by the structure factor is different from what we obtained in the analysis of the eigenvalues of $\mathcal{J}(\VEC{q})$ in Fig.~\ref{fig-2}e.
The eigenvectors corresponding to the lowest eigenvalue of $\mathcal{J}(\VEC{q})$ lead to single-Q states which do not necessarily have unit spin length in real space, and so cannot be directly compared to the results of the atomistic spin dynamics.
In addition, the in-plane anisotropy of the Pt/Mn/\textit{h}-BN structure also favors the stabilization of the commensurate spin texture observed in Fig.\ref{fig-3}a, and this was not included in the  $\mathcal{J}(\VEC{q})$ picture. 
The noncollinear state in Fig.~\ref{fig-3}a can be decomposed into three sublattices, labelled as Mn(1), Mn(2) and Mn(3).
The spin structure factors for each sublattice (Fig.~\ref{fig-3}b-d) reveal that each sublattice has a different dominant $\VEC{q}$, given by the green spots circled in black.
The magnetic arrangement of each sublattice can be approximately understood as a distorted spin spiral state, and we can obtain the main rotation axis about which the magnetic moments spiral by computing $\VEC{S}(\VEC{q})^*\times \VEC{S} (\VEC{q})$ from the dominant $\VEC{q}$ contribution to the spin structure factor.
These rotation axes are tilted from the $z$-axis by about 25$^\circ$ and are rotated 120$^\circ$ around the $z$-axis relative to each other.
Repeating the simulations starting from different initial random states leads to seemingly different magnetic configurations of this noncollinear 36-site magnetic state, but they can be verified to be equivalent by having the same properties of the spin structure factor just described.

The discovered noncollinear triple-Q state is noncoplanar, as shown in the three-dimensional image in Fig.~\ref{fig-3}a, so it will imprint an emergent magnetic field on the conduction electrons which defines the topological charge (see Method section and Fig.~S3). A finite integer charge confers topological properties to the underlying spin-texture, enhancing its stability by giving rise to a protecting topological barrier while inducing the topological Hall effect.
Within the 30$\times$30 simulation lattice, the topological charge reaches an integer value of $\sim$-300.
It is interesting to remark that, unlike the topological charge of skyrmions, which is governed by the DMI, 
the non-zero topological number of our triple-Q state is primarily contributed by the Heisenberg exchange interactions.
We would expect a rich set of nontrivial  Hall effects in this topologically nontrivial spin texture as an implication of its strong local noncoplanarity.

\begin{figure}
    \centering
    \includegraphics[width=1\linewidth]{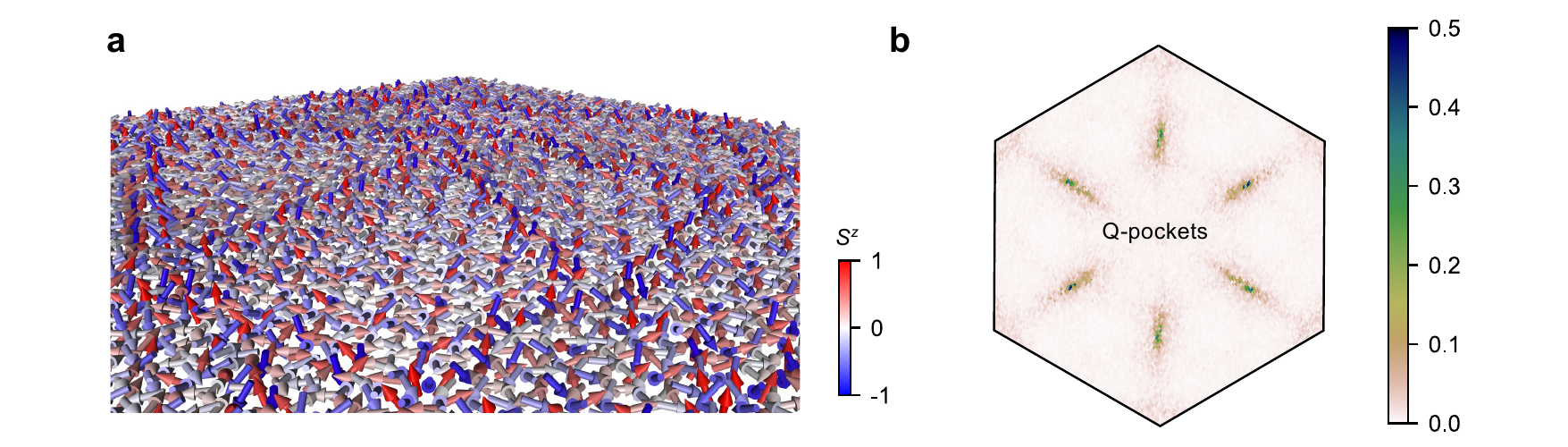}
    \caption{Image of the magnetic state and the structure factor with a 120$\times$120 simulation lattice. \textbf{a} Magnetic domains with an energy 0.18 meV/spin higher than the triple-Q state.
    \textbf{b} The structure factor of the magnetic state on the Pt/Mn/\textit{h}-BN kagome structure, relaxed within a 120$\times$120 simulation lattice at 0 K.
    }
    \label{fig-5}
\end{figure}

However, when we used a larger simulation lattice with 120$\times$120 unit cell, we could also observe a more complex spin state with an energy 0.18 meV/spin higher than the triple-Q state, as shown in Fig.~\ref{fig-5}a.
We can discern several magnetic domains, and the spin structure factor for this magnetic state with a slightly higher energy than the ground state is shown in Fig.~\ref{fig-5}b, where several Q-pockets can be observed.
Those Q-pockets lead to a superposition of a distribution of Q states with different periodicities residing in each pocket. Such features are reminiscent of the Q-space behavior of self-induced spin glass in the magnetic ground state of elemental crystalline Nd~\cite{doi:10.1126/science.aay6757}, which also arises from competing Heisenberg exchange interactions.
The small energy cost for creating such a superposition of different Q-states led us to investigate how the magnetic state of the system changes under thermal perturbations, with the results summarized in Supplementary Fig.~S4.
The main observation can be rationalized by considering the flat bands in Fig.~\ref{fig-2}f, which have a bandwidth of about 6 meV and corresponds to a temperature of 70 K.
Therefore, when the temperature in the atomistic spin dynamics simulations is increased past that threshold, the peaks of the spin structure factor melt away and we find a complex disordered magnetic state.
This demonstrates that the magnetic interactions embodied in the flat bands are indeed central in determining not only the ground state but also the thermal properties of this system.

\section*{Discussion}
We have previously proposed a method to form a monolayer 3\textit{d} transition metal kagome lattice using \textit{h}-BN as an aid and focused on its implications for ferromagnetic layers~\cite{Zhou2024}.
Inspired by this approach, here we investigated the frustrated magnetism of AFM Mn on a metastable kagome lattice realized in such a sandwiched structure, with Pt(111) as the substrate and \textit{h}-BN as the overlayer.
We adopted a two-pronged approach based on extracting all the relevant magnetic interactions from first-principles calculations combined with an exploration of the magnetic states through atomistic spin dynamics.
Our analysis shows that the geometrical frustration in the Mn kagome lattice leads to rather flat bands in the Fourier-transformed magnetic interactions, which signify the almost degeneracy of multiple magnetic states.
\textit{h}-BN breaks the symmetry of the ideal kagome lattice without significantly affecting the magnetic flat bands, resulting in a magnetic breathing kagome lattice.

Analysis of the Fourier transform of the magnetic interactions indicated a tendency for a noncollinear magnetic ground state, and atomistic spin dynamics simulations revealed that the ground state is indeed a specific type of triple-Q state that has not been previously reported, to the best of our knowledge.
We have fully characterized the properties of this triple-Q state by considering its real and reciprocal space structure, in particular exploiting the information encoded in its spin structure factor.
Additionally, we found that each of the three sublattices can be characterized as a distorted spin spiral state with a dominant $\VEC{q}$ vector, with the dominant $\VEC{q}$-vectors for different sublattices making 120$^\circ$ angles relative to each other, and their combination leading to the observed triple-Q state.

The most striking feature of this triple-Q state is its very strong local noncoplanarity, that we quantified by mapping its topological charge density.
Such a strong noncoplanarity will have a dramatic impact on the motion of conduction electrons through this magnetic state, and we anticipate highly nonlinear Hall effects as a consequence.
The relative simplicity of fabrication of such heterostructures in laboratories that are currently investigating 2D materials makes the experimental realization of our proposed system a very concrete possibility, and its unconventional magnetic and transport properties can be accessed by various techniques, most notably spin-polarized scanning tunneling microscopy.

Concerning the almost flat bands obtained for spin spirals, we also found a more complex spin states with a higher energy than triple-Q state, where the spin structure factor exhibits characteristics  reminiscent of  self-induced spin glass behavior. Notably, we observed that these flat bands also significantly influence the thermal properties of the system, which transitions into a disordered magnetic state when the temperature exceeds a certain threshold.

In conclusion, the atomic structure enabled by the kagomerization mechanism provides a novel platform for investigating frustration-related properties and complex noncollinear magnetic states.
Our findings offer valuable insights into the interplay between geometry and magnetism, paving the way for advancements in exploring novel spin textures.

\begin{methods}

\subsection{First-principles calculations.}
Optimization of the considered structures was performed using the density functional theory (DFT) approach implemented in the Quantum Espresso computational package~\cite{Giannozzi_2009}, including the van der Waals correction (DFT-D3~\cite{10.1063/1.3382344}).
We employed the projector augmented wave pseudopotentials from the pslibrary~\cite{DALCORSO2014337}, with the generalized gradient approximation of Perdew, Burke and Ernzerhof (PBE)~\cite{PhysRevLett.77.3865} as the exchange and correlation functional.
Convergence tests led to a plane-wave energy cut-off of 90 Ry and an 18$\times$18$\times$1 k-point grid.
We optimize the lattice constant of the HM substrates and use those values for the heterostructures with TM layers and with or without \textit{h}-BN.

The magnetic properties and magnetic interactions were computed using the all-electron full-potential scalar-relativistic Korringa-Kohn-Rostoker (KKR) Green function method~\cite{Papanikolaou_2002,bauer2014development,russmann_2022_7284739} including spin-orbit coupling self-consistently as implemented in the JuKKR computational package.
The angular momentum expansion of the Green function was truncated at $\ell_{\text{max}} = 3$ with a k-mesh of 18$\times$18$\times$1 points.
The energy integrations were performed including a Fermi-Dirac smearing of 502.78 K, and the Perdew-Wang 91 generalized gradient approximation  was employed~\cite{bauer2014development}. 
The Heisenberg exchange interactions and DM vectors were extracted using the infinitesimal rotation method ~\cite{LIECHTENSTEIN198765,PhysRevB.79.045209} with a finer k-mesh of 80$\times$80$\times$1 from the DLM ~\cite{BLGyorffy_1985} state. The DLM theory is built upon the identification of robust local magnetic moments at certain sites $\{n\}$, with orientations prescribed by unit vectors $\{\hat{\VEC{e}}_n\}$ under a specified probability distribution. The approach is based on the assumption of a time-scale separation between the evolution of  $\{\hat{\VEC{e}}_n\}$ in comparison to a rapidly adapting underlying electronic structure~\cite{BLGyorffy_1985}.
Following this adiabatic approximation, DFT calculations constrained to different magnetic configurations $\{\hat{\VEC{e}}_n\}$ can be performed to describe different states of magnetic order arising at finite temperatures. We extracted the magnetic interactions from fully disordered local moments whose orientations average to zero ($\braket{\hat{\VEC{e}}_n}=0\}$), which is the high-temperature paramagnetic state. 
Adopting the DLM state as the reference state instead of the usual ferromagnetic or simple collinear AFM states has the benefit of not biasing the obtained magnetic interactions by choosing a magnetic reference state that might be quite different from the actual ground state.

\subsection{Magnetic interactions and atomistic spin dynamics.}
We consider a classical extended Heisenberg Hamiltonian including Heisenberg exchange coupling ($J$), DMI ($\mathbf{D}$), the magnetic anisotropy energy ($K$), and Zeeman term ($\mathbf{B}$). All parameters were obtained from first-principles calculation. 
The \new{energy of the reference unit cell in this} spin-lattice model reads as follows ($|\VEC{S}|=1$):
\begin{equation}\label{eq:spin_model}
    E = -\sum_{n\mu\nu} J_{0\mu,n\nu}\,\mathbf{S}_{0\mu} \cdot \mathbf{S}_{n\nu}
    - \sum_{n\mu\nu} \mathbf{D}_{0\mu,n\nu}\cdot\left(\mathbf{S}_{0\mu} \times \mathbf{S}_{n\nu}\right) -\sum_{\mu} \mathbf{B}\cdot \mathbf{S}_{0\mu} - \sum_{\mu} K_{\mu} \left(S_{0\mu}^z\right)^2\;.
\end{equation}
where $n$ labels unit cells with 0 being the one including the origin. $\mu$ and $\nu$ label different magnetic sites within a unit cell.
The magnetic properties can be characterized by analysing the Fourier-transformed magnetic interactions, which in reciprocal space gives access to the magnetic ground state and the energy of spin spiral magnetic states:
\begin{align}
    J_{\mu\nu}(\VEC{q})&=\sum_n J_{0\mu,n\nu}e^{-\iu\VEC{q}\cdot(\VEC{R}_{0n}+\VEC{R}\mu\nu)}\\
    D_{\mu\nu,\alpha}(\VEC{q})&=\sum_n D_{0\mu,n\nu,\alpha}e^{-\iu\VEC{q}\cdot(\VEC{R}_{0n}+\VEC{R}\mu\nu)}
\end{align}
where  $\alpha=\{x,y,z\}$, $\VEC{R}_{0n}$ is a vector connecting unit cell 0 and $n$, while $\VEC{R}_{\mu\nu}$ is a vector connecting atoms $\mu$ and $\nu$ in the same unit cell.
The contribution to the energy from the two-site interactions can be expressed in reciprocal space with the following quantity,
\begin{equation}
    \mathcal{J}(\VEC{q})=\begin{pmatrix}
        J(\VEC{q})&D_z(\VEC{q})&-D_y(\VEC{q})\\
        -D_z(\VEC{q})&J(\VEC{q})&D_x(\VEC{q})\\
        D_y(\VEC{q})&-D_x(\VEC{q})&J(\VEC{q})
    \end{pmatrix} \;,
\end{equation}
which is a 9$\times$9 matrix (three magnetic sublattices $\times$ three spatial dimensions).
The eigenvectors of $\mathcal{J}(\VEC{q})$ provide a set of independent possible magnetic states, and the eigenvalues are their respective energies.

A definite answer regarding complex magnetic states is afforded through atomistic spin dynamic simulations using the Landau-Lifshitz-equation (LLG) as implemented in the Spirit code and performed using the Hamiltonian from Eq.~\ref{eq:spin_model}.
We employed periodic boundary conditions to model the extended two-dimensional system and used the simulated annealing method to navigate the intricate energy landscape and arrive at the ground state.
We started from a random spin state at 1000 K which we let equilibrate, then cooled the system in steps by reducing the temperature to half of its previous value and equilibrating again, until we reached below 1 K. During the annealing process, a 1 T external field was applied along $z$-axis to facilitate the formation of a single domain.
The spin structure factor is defined by $\VEC{S}(\VEC{q})=\frac{1}{N_c}\sum_i \VEC{S}_{i}\,e^{-\iu\VEC{q}\cdot\VEC{R}_i}$ with $N_c$ the numbers of spins in the simulation cell and $\VEC{S}_i$ is the magnetic moment orientation at position $\VEC{R}_i$.

\subsection{Topological charge.}

We know that magnetic skyrmions are characterized by an integer topological number corresponding to the total solid angle $n_{\rm sk}=\frac{1}{4\pi}\sum_{i,j,k}\Omega_{ijk}$ which can be related to the topological Hall effect.
The solid angle for the three spins $\Omega_{ijk}$ is related with the scalar chirality $\chi_{ijk}=\VEC{S}_i\cdot(\VEC{S}_j\times\VEC{S}_k)$ via  $\Omega_{ijk}=2\tan^{-1}(\frac{\chi_{ijk}}{|\VEC{S}_i||\VEC{S}_j||\VEC{S}_k|+\sum_{\rm cyclic}\VEC{S}_i\cdot\VEC{S}_j|\VEC{S}_k|})$~\cite{4121581}.
\new{We use the topological charge density defined by summing the solid angle for all triangles neighboring a given spin} to characterise the degree of noncoplanarity of the triple-Q state \new{on the magnetic breathing kagome lattice}. 
In \new{Supplementary} Fig.~S3b, we provide a schematic representation of the calculation method.
We use blue, green and orange to represent spins in three sublattices.
The colored rectangles (blue, green and orange) indicates the cells used for calculating the solid angle subtended by the central spin (matching the shadow's color) and its surrounding spins.
The cell with colored shading  encompasses six triangles, numbered in counterclockwise order.
We show the calculated topological charge density of each cell in Supplementary Fig.~S3c, the value of each cell is calculated by $n=\frac{1}{4\pi}\sum_{i,j,k}\Omega_{ijk}$, where $i$, $j$ and $k$ represents the three spins within each of the six triangles. We note that in \new{Supplementary} Fig.~S3c, there are dark blue and dark red points with values greater than 1, indicating that these cells exhibit very strong noncoplanarity.
\new{The topological charge is defined as the sum over the solid angles assigned to all the sites in the 30$\times$30 simulation lattice, and we obtain an integer value of -300.}
\new{However, this value can change depending on the details of the magnetic configuration, which we believe signals a breakdown in the lattice version of the topological charge calculation; this concept is mathematically well-defined only for continuum vector fields.}

\subsection{Data availability}
The data needed to evaluate the conclusions in the paper are present in the paper and the Supplementary Information.

\subsection{Code availability}
The codes employed for the simulations described within this work are open-source and can be obtained from the respective websites and/or repositories.
Quantum Espresso can be found at \cite{QEurl}, and the J\"ulich-developed codes JuKKR and Spirit can be found at \cite{JuKKRurl} and \cite{Spiriturl}, respectively.
\end{methods}

\begin{addendum}

\item

This work was supported by the National Key Research and Development Program of China (2022YFB4400200, W.Z.), the National Natural Science Foundation of China (No.T2394474, T2394470, W.Z.), the New Cornerstone Science Foundation through the XPLORER PRIZE (W.Z.), the Priority Programmes SPP 2244 “2D Materials Physics of van der Waals heterobilayer” (Project LO 1659/7-1, S.L.), SPP 2137 “Skyrmionics” (Project LO 1659/8-1, S.L.) of the Deutsche Forschungsgemeinschaft (DFG), 
 CoSeC and the Computational Science Centre for Research Communities (CCP9, M.d.S.D.).   Simulations were performed with computing resources granted by RWTH Aachen University under project p0020362 and JARA on the supercomputer JURECA~\cite{jureca} at Forschungszentrum J\"ulich.

\item[Author contributions]
S.L. initiated, designed and supervised the project. H.Z. performed the simulations with support and supervision from M.d.S.D, W.Z. and S.L. H.Z., M.d.S.D., S.B., H.L., Y.Z., W.Z. and S.L. discussed the results. H.Z., M.d.S.D and S.L. wrote the manuscript to which all co-authors contributed.
\item[Competing interests]
The authors declare no competing interests.

\end{addendum}

\clearpage
\section*{References}

\bibliography{references}

\begin{thebibliography}{10}
\expandafter\ifx\csname url\endcsname\relax
  \def\url#1{\texttt{#1}}\fi
\expandafter\ifx\csname urlprefix\endcsname\relax\def\urlprefix{URL }\fi
\providecommand{\bibinfo}[2]{#2}
\providecommand{\eprint}[2][]{\url{#2}}

\bibitem{PhysRevB.81.014406}
\bibinfo{author}{Andreanov, A.}, \bibinfo{author}{Chalker, J.~T.},
  \bibinfo{author}{Saunders, T.~E.} \& \bibinfo{author}{Sherrington, D.}
\newblock \bibinfo{title}{Spin-glass transition in geometrically frustrated
  antiferromagnets with weak disorder}.
\newblock \emph{\bibinfo{journal}{Phys. Rev. B}} \textbf{\bibinfo{volume}{81}},
  \bibinfo{pages}{014406} (\bibinfo{year}{2010}).
\newblock \urlprefix\url{https://link.aps.org/doi/10.1103/PhysRevB.81.014406}.

\bibitem{doi:10.1143/JPSJ.79.011003}
\bibinfo{author}{Nakatsuji, S.}, \bibinfo{author}{Nambu, Y.} \&
  \bibinfo{author}{Onoda, S.}
\newblock \bibinfo{title}{Novel geometrical frustration effects in the
  two-dimensional triangular-lattice antiferromagnet
  {NiGa\textsubscript{2}S\textsubscript{4}} and related compounds}.
\newblock \emph{\bibinfo{journal}{J. Phys. Soc. Japan}}
  \textbf{\bibinfo{volume}{79}}, \bibinfo{pages}{011003}
  (\bibinfo{year}{2010}).
\newblock \urlprefix\url{https://doi.org/10.1143/JPSJ.79.011003}.
\newblock \eprint{https://doi.org/10.1143/JPSJ.79.011003}.

\bibitem{PhysRevB.81.094407}
\bibinfo{author}{Igoshev, P.~A.}, \bibinfo{author}{Timirgazin, M.~A.},
  \bibinfo{author}{Katanin, A.~A.}, \bibinfo{author}{Arzhnikov, A.~K.} \&
  \bibinfo{author}{Irkhin, V.~Y.}
\newblock \bibinfo{title}{Incommensurate magnetic order and phase separation in
  the two-dimensional {Hubbard} model with nearest- and next-nearest-neighbor
  hopping}.
\newblock \emph{\bibinfo{journal}{Phys. Rev. B}} \textbf{\bibinfo{volume}{81}},
  \bibinfo{pages}{094407} (\bibinfo{year}{2010}).
\newblock \urlprefix\url{https://link.aps.org/doi/10.1103/PhysRevB.81.094407}.

\bibitem{PhysRevB.103.054422}
\bibinfo{author}{Hayami, S.} \& \bibinfo{author}{Motome, Y.}
\newblock \bibinfo{title}{Noncoplanar multiple-$q$ spin textures by itinerant
  frustration: Effects of single-ion anisotropy and bond-dependent anisotropy}.
\newblock \emph{\bibinfo{journal}{Phys. Rev. B}}
  \textbf{\bibinfo{volume}{103}}, \bibinfo{pages}{054422}
  (\bibinfo{year}{2021}).
\newblock \urlprefix\url{https://link.aps.org/doi/10.1103/PhysRevB.103.054422}.

\bibitem{doi:10.1126/sciadv.aau3402}
\bibinfo{author}{Takagi, R.} \emph{et~al.}
\newblock \bibinfo{title}{Multiple-\textit{q} noncollinear magnetism in an
  itinerant hexagonal magnet}.
\newblock \emph{\bibinfo{journal}{Sci. Adv.}} \textbf{\bibinfo{volume}{4}},
  \bibinfo{pages}{eaau3402} (\bibinfo{year}{2018}).
\newblock
  \urlprefix\url{https://www.science.org/doi/abs/10.1126/sciadv.aau3402}.
\newblock \eprint{https://www.science.org/doi/pdf/10.1126/sciadv.aau3402}.

\bibitem{PhysRevLett.86.1106}
\bibinfo{author}{Kurz, P.}, \bibinfo{author}{Bihlmayer, G.},
  \bibinfo{author}{Hirai, K.} \& \bibinfo{author}{Bl\"ugel, S.}
\newblock \bibinfo{title}{Three-dimensional spin structure on a two-dimensional
  lattice: {Mn$/$Cu}(111)}.
\newblock \emph{\bibinfo{journal}{Phys. Rev. Lett.}}
  \textbf{\bibinfo{volume}{86}}, \bibinfo{pages}{1106--1109}
  (\bibinfo{year}{2001}).
\newblock \urlprefix\url{https://link.aps.org/doi/10.1103/PhysRevLett.86.1106}.

\bibitem{PhysRevLett.124.227203}
\bibinfo{author}{Spethmann, J.} \emph{et~al.}
\newblock \bibinfo{title}{Discovery of magnetic single- and triple-$\mathbf{q}$
  states in {Mn}/{Re}(0001)}.
\newblock \emph{\bibinfo{journal}{Phys. Rev. Lett.}}
  \textbf{\bibinfo{volume}{124}}, \bibinfo{pages}{227203}
  (\bibinfo{year}{2020}).
\newblock
  \urlprefix\url{https://link.aps.org/doi/10.1103/PhysRevLett.124.227203}.

\bibitem{PhysRevB.95.224424}
\bibinfo{author}{Hayami, S.}, \bibinfo{author}{Ozawa, R.} \&
  \bibinfo{author}{Motome, Y.}
\newblock \bibinfo{title}{Effective bilinear-biquadratic model for noncoplanar
  ordering in itinerant magnets}.
\newblock \emph{\bibinfo{journal}{Phys. Rev. B}} \textbf{\bibinfo{volume}{95}},
  \bibinfo{pages}{224424} (\bibinfo{year}{2017}).
\newblock \urlprefix\url{https://link.aps.org/doi/10.1103/PhysRevB.95.224424}.

\bibitem{Heinze2011}
\bibinfo{author}{Heinze, S.} \emph{et~al.}
\newblock \bibinfo{title}{Spontaneous atomic-scale magnetic skyrmion lattice in
  two dimensions}.
\newblock \emph{\bibinfo{journal}{Nat. Phys.}} \textbf{\bibinfo{volume}{7}},
  \bibinfo{pages}{713--718} (\bibinfo{year}{2011}).
\newblock \urlprefix\url{https://doi.org/10.1038/nphys2045}.

\bibitem{PhysRevB.92.020401}
\bibinfo{author}{Hoffmann, M.} \emph{et~al.}
\newblock \bibinfo{title}{Topological orbital magnetization and emergent {Hall}
  effect of an atomic-scale spin lattice at a surface}.
\newblock \emph{\bibinfo{journal}{Phys. Rev. B}} \textbf{\bibinfo{volume}{92}},
  \bibinfo{pages}{020401} (\bibinfo{year}{2015}).
\newblock \urlprefix\url{https://link.aps.org/doi/10.1103/PhysRevB.92.020401}.

\bibitem{PhysRevLett.95.057205}
\bibinfo{author}{Katsura, H.}, \bibinfo{author}{Nagaosa, N.} \&
  \bibinfo{author}{Balatsky, A.~V.}
\newblock \bibinfo{title}{Spin current and magnetoelectric effect in
  noncollinear magnets}.
\newblock \emph{\bibinfo{journal}{Phys. Rev. Lett.}}
  \textbf{\bibinfo{volume}{95}}, \bibinfo{pages}{057205}
  (\bibinfo{year}{2005}).
\newblock
  \urlprefix\url{https://link.aps.org/doi/10.1103/PhysRevLett.95.057205}.

\bibitem{PhysRevLett.113.196602}
\bibinfo{author}{Zhang, W.} \emph{et~al.}
\newblock \bibinfo{title}{Spin {H}all effects in metallic antiferromagnets}.
\newblock \emph{\bibinfo{journal}{Phys. Rev. Lett.}}
  \textbf{\bibinfo{volume}{113}}, \bibinfo{pages}{196602}
  (\bibinfo{year}{2014}).
\newblock
  \urlprefix\url{https://link.aps.org/doi/10.1103/PhysRevLett.113.196602}.

\bibitem{PhysRevB.45.13544}
\bibinfo{author}{Loss, D.} \& \bibinfo{author}{Goldbart, P.~M.}
\newblock \bibinfo{title}{Persistent currents from {Berry's} phase in
  mesoscopic systems}.
\newblock \emph{\bibinfo{journal}{Phys. Rev. B}} \textbf{\bibinfo{volume}{45}},
  \bibinfo{pages}{13544--13561} (\bibinfo{year}{1992}).
\newblock \urlprefix\url{https://link.aps.org/doi/10.1103/PhysRevB.45.13544}.

\bibitem{PhysRevLett.83.3737}
\bibinfo{author}{Ye, J.} \emph{et~al.}
\newblock \bibinfo{title}{Berry phase theory of the anomalous {H}all effect:
  {A}pplication to colossal magnetoresistance manganites}.
\newblock \emph{\bibinfo{journal}{Phys. Rev. Lett.}}
  \textbf{\bibinfo{volume}{83}}, \bibinfo{pages}{3737--3740}
  (\bibinfo{year}{1999}).
\newblock \urlprefix\url{https://link.aps.org/doi/10.1103/PhysRevLett.83.3737}.

\bibitem{PhysRevB.62.R6065}
\bibinfo{author}{Ohgushi, K.}, \bibinfo{author}{Murakami, S.} \&
  \bibinfo{author}{Nagaosa, N.}
\newblock \bibinfo{title}{Spin anisotropy and quantum {Hall} effect in the
  kagom\'e lattice: {Chiral} spin state based on a ferromagnet}.
\newblock \emph{\bibinfo{journal}{Phys. Rev. B}} \textbf{\bibinfo{volume}{62}},
  \bibinfo{pages}{R6065--R6068} (\bibinfo{year}{2000}).
\newblock \urlprefix\url{https://link.aps.org/doi/10.1103/PhysRevB.62.R6065}.

\bibitem{Surgers2014}
\bibinfo{author}{S{\"u}rgers, C.}, \bibinfo{author}{Fischer, G.},
  \bibinfo{author}{Winkel, P.} \& \bibinfo{author}{L{\"o}hneysen, H.~v.}
\newblock \bibinfo{title}{Large topological {Hall} effect in the non-collinear
  phase of an antiferromagnet}.
\newblock \emph{\bibinfo{journal}{Nat. Commun.}} \textbf{\bibinfo{volume}{5}},
  \bibinfo{pages}{3400} (\bibinfo{year}{2014}).
\newblock \urlprefix\url{https://doi.org/10.1038/ncomms4400}.

\bibitem{Xu2024}
\bibinfo{author}{Xu, S.} \emph{et~al.}
\newblock \bibinfo{title}{Universal scaling law for chiral antiferromagnetism}.
\newblock \emph{\bibinfo{journal}{Nat. Commun.}} \textbf{\bibinfo{volume}{15}},
  \bibinfo{pages}{3717} (\bibinfo{year}{2024}).
\newblock \urlprefix\url{https://doi.org/10.1038/s41467-024-46325-5}.

\bibitem{PhysRevLett.126.147203}
\bibinfo{author}{Bouaziz, J.}, \bibinfo{author}{Ishida, H.},
  \bibinfo{author}{Lounis, S.} \& \bibinfo{author}{Bl\"ugel, S.}
\newblock \bibinfo{title}{Transverse transport in two-dimensional relativistic
  systems with nontrivial spin textures}.
\newblock \emph{\bibinfo{journal}{Phys. Rev. Lett.}}
  \textbf{\bibinfo{volume}{126}}, \bibinfo{pages}{147203}
  (\bibinfo{year}{2021}).
\newblock
  \urlprefix\url{https://link.aps.org/doi/10.1103/PhysRevLett.126.147203}.

\bibitem{Ishizuka2020}
\bibinfo{author}{Ishizuka, H.} \& \bibinfo{author}{Nagaosa, N.}
\newblock \bibinfo{title}{Anomalous electrical magnetochiral effect by chiral
  spin-cluster scattering}.
\newblock \emph{\bibinfo{journal}{Nat. Commun.}} \textbf{\bibinfo{volume}{11}},
  \bibinfo{pages}{2986} (\bibinfo{year}{2020}).
\newblock \urlprefix\url{https://doi.org/10.1038/s41467-020-16751-2}.

\bibitem{PhysRevB.101.220403}
\bibinfo{author}{Hayami, S.}, \bibinfo{author}{Yanagi, Y.} \&
  \bibinfo{author}{Kusunose, H.}
\newblock \bibinfo{title}{Spontaneous antisymmetric spin splitting in
  noncollinear antiferromagnets without spin-orbit coupling}.
\newblock \emph{\bibinfo{journal}{Phys. Rev. B}}
  \textbf{\bibinfo{volume}{101}}, \bibinfo{pages}{220403}
  (\bibinfo{year}{2020}).
\newblock \urlprefix\url{https://link.aps.org/doi/10.1103/PhysRevB.101.220403}.

\bibitem{Crum2015}
\bibinfo{author}{Crum, D.~M.} \emph{et~al.}
\newblock \bibinfo{title}{Perpendicular reading of single confined magnetic
  skyrmions}.
\newblock \emph{\bibinfo{journal}{Nat. Commun.}} \textbf{\bibinfo{volume}{6}},
  \bibinfo{pages}{8541} (\bibinfo{year}{2015}).
\newblock \urlprefix\url{https://doi.org/10.1038/ncomms9541}.

\bibitem{LimaFernandes2022}
\bibinfo{author}{Lima~Fernandes, I.}, \bibinfo{author}{Bl{\"u}gel, S.} \&
  \bibinfo{author}{Lounis, S.}
\newblock \bibinfo{title}{Spin-orbit enabled all-electrical readout of chiral
  spin-textures}.
\newblock \emph{\bibinfo{journal}{Nat. Commun.}} \textbf{\bibinfo{volume}{13}},
  \bibinfo{pages}{1576} (\bibinfo{year}{2022}).
\newblock \urlprefix\url{https://doi.org/10.1038/s41467-022-29237-0}.

\bibitem{dosSantosDias2016}
\bibinfo{author}{dos Santos~Dias, M.}, \bibinfo{author}{Bouaziz, J.},
  \bibinfo{author}{Bouhassoune, M.}, \bibinfo{author}{Bl{\"u}gel, S.} \&
  \bibinfo{author}{Lounis, S.}
\newblock \bibinfo{title}{Chirality-driven orbital magnetic moments as a new
  probe for topological magnetic structures}.
\newblock \emph{\bibinfo{journal}{Nat. Commun.}} \textbf{\bibinfo{volume}{7}},
  \bibinfo{pages}{13613} (\bibinfo{year}{2016}).
\newblock \urlprefix\url{https://doi.org/10.1038/ncomms13613}.

\bibitem{PhysRevB.94.121114}
\bibinfo{author}{Hanke, J.-P.} \emph{et~al.}
\newblock \bibinfo{title}{Role of berry phase theory for describing orbital
  magnetism: {From} magnetic heterostructures to topological orbital
  ferromagnets}.
\newblock \emph{\bibinfo{journal}{Phys. Rev. B}} \textbf{\bibinfo{volume}{94}},
  \bibinfo{pages}{121114} (\bibinfo{year}{2016}).
\newblock \urlprefix\url{https://link.aps.org/doi/10.1103/PhysRevB.94.121114}.

\bibitem{PhysRevB.98.125420}
\bibinfo{author}{Bouaziz, J.}, \bibinfo{author}{Dias, M. d.~S.},
  \bibinfo{author}{Guimar\~aes, F. S.~M.}, \bibinfo{author}{Bl\"ugel, S.} \&
  \bibinfo{author}{Lounis, S.}
\newblock \bibinfo{title}{Impurity-induced orbital magnetization in a {Rashba}
  electron gas}.
\newblock \emph{\bibinfo{journal}{Phys. Rev. B}} \textbf{\bibinfo{volume}{98}},
  \bibinfo{pages}{125420} (\bibinfo{year}{2018}).
\newblock \urlprefix\url{https://link.aps.org/doi/10.1103/PhysRevB.98.125420}.

\bibitem{PhysRevB.98.094428}
\bibinfo{author}{Brinker, S.}, \bibinfo{author}{dos Santos~Dias, M.} \&
  \bibinfo{author}{Lounis, S.}
\newblock \bibinfo{title}{Interatomic orbital magnetism: The case of $3d$
  adatoms deposited on the {Pt(111)} surface}.
\newblock \emph{\bibinfo{journal}{Phys. Rev. B}} \textbf{\bibinfo{volume}{98}},
  \bibinfo{pages}{094428} (\bibinfo{year}{2018}).
\newblock \urlprefix\url{https://link.aps.org/doi/10.1103/PhysRevB.98.094428}.

\bibitem{Grytsiuk2020}
\bibinfo{author}{Grytsiuk, S.} \emph{et~al.}
\newblock \bibinfo{title}{Topological--chiral magnetic interactions driven by
  emergent orbital magnetism}.
\newblock \emph{\bibinfo{journal}{Nat. Commun.}} \textbf{\bibinfo{volume}{11}},
  \bibinfo{pages}{511} (\bibinfo{year}{2020}).
\newblock \urlprefix\url{https://doi.org/10.1038/s41467-019-14030-3}.

\bibitem{WORTMANN200257}
\bibinfo{author}{Wortmann, D.} \emph{et~al.}
\newblock \bibinfo{title}{Resolving noncollinear magnetism by spin-polarized
  scanning tunneling microscopy}.
\newblock \emph{\bibinfo{journal}{J. Magn. Magn. Mater.}}
  \textbf{\bibinfo{volume}{240}}, \bibinfo{pages}{57--63}
  (\bibinfo{year}{2002}).
\newblock
  \urlprefix\url{https://www.sciencedirect.com/science/article/pii/S0304885301007387}.
\newblock \bibinfo{note}{4th International Symposium on Metallic Multilayers}.

\bibitem{PhysRevB.108.L180411}
\bibinfo{author}{Nickel, F.} \emph{et~al.}
\newblock \bibinfo{title}{Coupling of the triple-$\mathrm{q}$ state to the
  atomic lattice by anisotropic symmetric exchange}.
\newblock \emph{\bibinfo{journal}{Phys. Rev. B}}
  \textbf{\bibinfo{volume}{108}}, \bibinfo{pages}{L180411}
  (\bibinfo{year}{2023}).
\newblock
  \urlprefix\url{https://link.aps.org/doi/10.1103/PhysRevB.108.L180411}.

\bibitem{PhysRevB.45.7536}
\bibinfo{author}{Huse, D.~A.} \& \bibinfo{author}{Rutenberg, A.~D.}
\newblock \bibinfo{title}{Classical antiferromagnets on the kagom\'e lattice}.
\newblock \emph{\bibinfo{journal}{Phys. Rev. B}} \textbf{\bibinfo{volume}{45}},
  \bibinfo{pages}{7536--7539} (\bibinfo{year}{1992}).
\newblock \urlprefix\url{https://link.aps.org/doi/10.1103/PhysRevB.45.7536}.

\bibitem{PhysRevB.78.094423}
\bibinfo{author}{Zhitomirsky, M.~E.}
\newblock \bibinfo{title}{Octupolar ordering of classical kagome
  antiferromagnets in two and three dimensions}.
\newblock \emph{\bibinfo{journal}{Phys. Rev. B}} \textbf{\bibinfo{volume}{78}},
  \bibinfo{pages}{094423} (\bibinfo{year}{2008}).
\newblock \urlprefix\url{https://link.aps.org/doi/10.1103/PhysRevB.78.094423}.

\bibitem{doi:10.1126/science.1201080}
\bibinfo{author}{Yan, S.}, \bibinfo{author}{Huse, D.~A.} \&
  \bibinfo{author}{White, S.~R.}
\newblock \bibinfo{title}{Spin-liquid ground state of the \textit{S} = 1/2
  kagome {H}eisenberg antiferromagnet}.
\newblock \emph{\bibinfo{journal}{Science}} \textbf{\bibinfo{volume}{332}},
  \bibinfo{pages}{1173--1176} (\bibinfo{year}{2011}).
\newblock
  \urlprefix\url{https://www.science.org/doi/abs/10.1126/science.1201080}.
\newblock \eprint{https://www.science.org/doi/pdf/10.1126/science.1201080}.

\bibitem{PhysRevB.83.212401}
\bibinfo{author}{L\"auchli, A.~M.}, \bibinfo{author}{Sudan, J.} \&
  \bibinfo{author}{S\o{}rensen, E.~S.}
\newblock \bibinfo{title}{Ground-state energy and spin gap of
  spin-$\frac{1}{2}$ kagom\'e-heisenberg antiferromagnetic clusters:
  {Large}-scale exact diagonalization results}.
\newblock \emph{\bibinfo{journal}{Phys. Rev. B}} \textbf{\bibinfo{volume}{83}},
  \bibinfo{pages}{212401} (\bibinfo{year}{2011}).
\newblock \urlprefix\url{https://link.aps.org/doi/10.1103/PhysRevB.83.212401}.

\bibitem{Jiang2012}
\bibinfo{author}{Jiang, H.-C.}, \bibinfo{author}{Wang, Z.} \&
  \bibinfo{author}{Balents, L.}
\newblock \bibinfo{title}{Identifying topological order by entanglement
  entropy}.
\newblock \emph{\bibinfo{journal}{Nat. Phys.}} \textbf{\bibinfo{volume}{8}},
  \bibinfo{pages}{902--905} (\bibinfo{year}{2012}).
\newblock \urlprefix\url{https://doi.org/10.1038/nphys2465}.

\bibitem{PhysRevB.76.180407}
\bibinfo{author}{Singh, R. R.~P.} \& \bibinfo{author}{Huse, D.~A.}
\newblock \bibinfo{title}{Ground state of the spin-1/2 kagome-lattice
  {H}eisenberg antiferromagnet}.
\newblock \emph{\bibinfo{journal}{Phys. Rev. B}} \textbf{\bibinfo{volume}{76}},
  \bibinfo{pages}{180407} (\bibinfo{year}{2007}).
\newblock \urlprefix\url{https://link.aps.org/doi/10.1103/PhysRevB.76.180407}.

\bibitem{Zhou2024}
\bibinfo{author}{Zhou, H.}, \bibinfo{author}{dos Santos~Dias, M.},
  \bibinfo{author}{Zhang, Y.}, \bibinfo{author}{Zhao, W.} \&
  \bibinfo{author}{Lounis, S.}
\newblock \bibinfo{title}{Kagomerization of transition metal monolayers induced
  by two-dimensional hexagonal boron nitride}.
\newblock \emph{\bibinfo{journal}{Nat. Commun.}} \textbf{\bibinfo{volume}{15}},
  \bibinfo{pages}{4854} (\bibinfo{year}{2024}).
\newblock \urlprefix\url{https://doi.org/10.1038/s41467-024-48973-z}.

\bibitem{PhysRevB.105.L100407}
\bibinfo{author}{Aoyama, K.} \& \bibinfo{author}{Kawamura, H.}
\newblock \bibinfo{title}{Emergent skyrmion-based chiral order in zero-field
  heisenberg antiferromagnets on the breathing kagome lattice}.
\newblock \emph{\bibinfo{journal}{Phys. Rev. B}}
  \textbf{\bibinfo{volume}{105}}, \bibinfo{pages}{L100407}
  (\bibinfo{year}{2022}).
\newblock
  \urlprefix\url{https://link.aps.org/doi/10.1103/PhysRevB.105.L100407}.

\bibitem{doi:10.1021/acs.jpclett.3c02419}
\bibinfo{author}{Aldarawsheh, A.}, \bibinfo{author}{Sallermann, M.},
  \bibinfo{author}{Abusaa, M.} \& \bibinfo{author}{Lounis, S.}
\newblock \bibinfo{title}{Intrinsic {N}éel antiferromagnetic multimeronic spin
  textures in ultrathin films}.
\newblock \emph{\bibinfo{journal}{J. Phys. Chem. Lett}}
  \textbf{\bibinfo{volume}{14}}, \bibinfo{pages}{8970--8978}
  (\bibinfo{year}{2023}).
\newblock \urlprefix\url{https://doi.org/10.1021/acs.jpclett.3c02419}.
\newblock \bibinfo{note}{PMID: 37773009},
  \eprint{https://doi.org/10.1021/acs.jpclett.3c02419}.

\bibitem{doi:10.1126/science.aay6757}
\bibinfo{author}{Kamber, U.} \emph{et~al.}
\newblock \bibinfo{title}{Self-induced spin glass state in elemental and
  crystalline neodymium}.
\newblock \emph{\bibinfo{journal}{Science}} \textbf{\bibinfo{volume}{368}},
  \bibinfo{pages}{eaay6757} (\bibinfo{year}{2020}).
\newblock
  \urlprefix\url{https://www.science.org/doi/abs/10.1126/science.aay6757}.
\newblock \eprint{https://www.science.org/doi/pdf/10.1126/science.aay6757}.

\bibitem{Giannozzi_2009}
\bibinfo{author}{Giannozzi, P.} \emph{et~al.}
\newblock \bibinfo{title}{{QUANTUM ESPRESSO}: a modular and open-source
  software project for quantum simulations of materials}.
\newblock \emph{\bibinfo{journal}{J. Phys.: Condens. Matter}}
  \textbf{\bibinfo{volume}{21}}, \bibinfo{pages}{395502}
  (\bibinfo{year}{2009}).
\newblock \urlprefix\url{https://dx.doi.org/10.1088/0953-8984/21/39/395502}.

\bibitem{10.1063/1.3382344}
\bibinfo{author}{Grimme, S.}, \bibinfo{author}{Antony, J.},
  \bibinfo{author}{Ehrlich, S.} \& \bibinfo{author}{Krieg, H.}
\newblock \bibinfo{title}{A consistent and accurate \textit{ab initio}
  parametrization of density functional dispersion correction ({DFT-D}) for the
  94 elements {H}-{P}u}.
\newblock \emph{\bibinfo{journal}{J. Chem. Phys.}}
  \textbf{\bibinfo{volume}{132}}, \bibinfo{pages}{154104}
  (\bibinfo{year}{2010}).
\newblock \urlprefix\url{https://doi.org/10.1063/1.3382344}.
\newblock
  \eprint{https://pubs.aip.org/aip/jcp/article-pdf/doi/10.1063/1.3382344/15684000/154104\_1\_online.pdf}.

\bibitem{DALCORSO2014337}
\bibinfo{author}{{Dal Corso}, A.}
\newblock \bibinfo{title}{Pseudopotentials periodic table: From {H} to {P}u}.
\newblock \emph{\bibinfo{journal}{Comp. Mater. Sci.}}
  \textbf{\bibinfo{volume}{95}}, \bibinfo{pages}{337--350}
  (\bibinfo{year}{2014}).
\newblock
  \urlprefix\url{https://www.sciencedirect.com/science/article/pii/S0927025614005187}.

\bibitem{PhysRevLett.77.3865}
\bibinfo{author}{Perdew, J.~P.}, \bibinfo{author}{Burke, K.} \&
  \bibinfo{author}{Ernzerhof, M.}
\newblock \bibinfo{title}{Generalized gradient approximation made simple}.
\newblock \emph{\bibinfo{journal}{Phys. Rev. Lett.}}
  \textbf{\bibinfo{volume}{77}}, \bibinfo{pages}{3865--3868}
  (\bibinfo{year}{1996}).
\newblock \urlprefix\url{https://link.aps.org/doi/10.1103/PhysRevLett.77.3865}.

\bibitem{Papanikolaou_2002}
\bibinfo{author}{Papanikolaou, N.}, \bibinfo{author}{Zeller, R.} \&
  \bibinfo{author}{Dederichs, P.~H.}
\newblock \bibinfo{title}{Conceptual improvements of the {KKR} method}.
\newblock \emph{\bibinfo{journal}{J. Phys.: Condens. Matter}}
  \textbf{\bibinfo{volume}{14}}, \bibinfo{pages}{2799} (\bibinfo{year}{2002}).
\newblock \urlprefix\url{https://dx.doi.org/10.1088/0953-8984/14/11/304}.

\bibitem{bauer2014development}
\bibinfo{author}{Bauer, D. S.~G.}
\newblock \emph{\bibinfo{title}{Development of a relativistic full-potential
  first-principles multiple scattering Green function method applied to complex
  magnetic textures of nanostructures at surfaces}}.
\newblock Ph.D. thesis, \bibinfo{school}{Aachen, Techn. Hochsch., Diss., 2013}
  (\bibinfo{year}{2014}).

\bibitem{russmann_2022_7284739}
\bibinfo{author}{Rüßmann, P.} \emph{et~al.}
\newblock \bibinfo{title}{Judftteam/jukkr: v3.6} (\bibinfo{year}{2022}).
\newblock \urlprefix\url{https://doi.org/10.5281/zenodo.7284739}.

\bibitem{LIECHTENSTEIN198765}
\bibinfo{author}{Liechtenstein, A.}, \bibinfo{author}{Katsnelson, M.},
  \bibinfo{author}{Antropov, V.} \& \bibinfo{author}{Gubanov, V.}
\newblock \bibinfo{title}{Local spin density functional approach to the theory
  of exchange interactions in ferromagnetic metals and alloys}.
\newblock \emph{\bibinfo{journal}{J. Magn. Magn. Mater.}}
  \textbf{\bibinfo{volume}{67}}, \bibinfo{pages}{65--74}
  (\bibinfo{year}{1987}).
\newblock
  \urlprefix\url{https://www.sciencedirect.com/science/article/pii/0304885387907219}.

\bibitem{PhysRevB.79.045209}
\bibinfo{author}{Ebert, H.} \& \bibinfo{author}{Mankovsky, S.}
\newblock \bibinfo{title}{Anisotropic exchange coupling in diluted magnetic
  semiconductors: \textit{Ab initio} spin-density functional theory}.
\newblock \emph{\bibinfo{journal}{Phys. Rev. B}} \textbf{\bibinfo{volume}{79}},
  \bibinfo{pages}{045209} (\bibinfo{year}{2009}).
\newblock \urlprefix\url{https://link.aps.org/doi/10.1103/PhysRevB.79.045209}.

\bibitem{BLGyorffy_1985}
\bibinfo{author}{Gyorffy, B.~L.}, \bibinfo{author}{Pindor, A.~J.},
  \bibinfo{author}{Staunton, J.}, \bibinfo{author}{Stocks, G.~M.} \&
  \bibinfo{author}{Winter, H.}
\newblock \bibinfo{title}{A first-principles theory of ferromagnetic phase
  transitions in metals}.
\newblock \emph{\bibinfo{journal}{Journal of Physics F: Metal Physics}}
  \textbf{\bibinfo{volume}{15}}, \bibinfo{pages}{1337} (\bibinfo{year}{1985}).
\newblock \urlprefix\url{https://dx.doi.org/10.1088/0305-4608/15/6/018}.

\bibitem{4121581}
\bibinfo{author}{Van~Oosterom, A.} \& \bibinfo{author}{Strackee, J.}
\newblock \bibinfo{title}{The solid angle of a plane triangle}.
\newblock \emph{\bibinfo{journal}{IEEE. Trans. Biomed. Eng.}}
  \textbf{\bibinfo{volume}{BME-30}}, \bibinfo{pages}{125--126}
  (\bibinfo{year}{1983}).

\bibitem{QEurl}
\bibinfo{note}{Website: \url{https://www.quantum-espresso.org}}.

\bibitem{JuKKRurl}
\bibinfo{note}{Repository: \url{https://github.com/JuDFTteam/JuKKR}}.

\bibitem{Spiriturl}
\bibinfo{note}{Website: \url{https://spirit-code.github.io}}.

\bibitem{jureca}
\bibinfo{author}{{J\"{u}lich Supercomputing Centre}}.
\newblock \bibinfo{title}{{JURECA}: Data centric and booster modules
  implementing the modular supercomputing architecture at {J}\"{u}lich
  supercomputing centre}.
\newblock \emph{\bibinfo{journal}{J. large-scale Res. facilities}}
  \textbf{\bibinfo{volume}{7}}, \bibinfo{pages}{A182} (\bibinfo{year}{2021}).
\newblock \urlprefix\url{http://dx.doi.org/10.17815/jlsrf-7-182}.

\end{thebibliography}
\newpage

\end{document}